\documentclass[12pt]{JHEP3}

\newcommand{\bea}{\begin{eqnarray}}
\newcommand{\beal}[1]{\begin{eqnarray}\label{#1}}
\newcommand{\eea}{\end{eqnarray}} 
\newcommand{\be}{\begin{equation}} 
\newcommand{\bel}[1]{\begin{equation}\label{#1}}
\newcommand{\ee}{\end{equation}}

\newcommand{\bit}{\begin{itemize}}
\newcommand{\eit}{\end{itemize}}
\newcommand{\ben}{\begin{enumerate}}
\newcommand{\een}{\end{enumerate}}

\newcommand{\req}[1]{(\ref{#1})}

\def\d{\partial}
\def\half{\frac{1}{2}}

\def\quart{\frac{1}{4}}

\def\mps{M_P^2}

\def\alp{\leavevmode\ifmmode {\alpha^\prime} \else ${\alpha^\prime}$ \fi}

\title{On the Slow Roll Expansion for Brane Inflation}

\preprint{}

\author{Micha\l\ Spali\'nski \footnote{Email: mspal@fuw.edu.pl} \\
So\l tan Institute for Nuclear Studies\\
ul. Ho\.za 69, 
00-681 Warszawa, Polska.
}

\abstract{One possibility for identifying the inflaton in the framework of
string theory is that it is a $D$-brane modulus. This option involves a
specific, non-canonical form of the kinetic energy -- the Dirac-Born-Infeld
action.  This note investigates the applicability of the slow roll
approximation in inflationary models of this type.  To this end the slow
roll expansion of Liddle, Parsons and Barrow is derived for the case of the
DBI action. The resulting slow roll conditions augment the standard ones
valid in the case of canonical kinetic terms. It is also shown that in DBI
models inflation does not require that the potential dominate the energy
density.}


\begin{document}

\section{Introduction}

The success of inflationary cosmology makes it important 
to look for its realization in string theory. In particular, it is important
to understand what possibilities exist for identifying the degrees of freedom
which could lead to an inflationary effective field theory at energies 
below the Planck scale. For a long time geometric closed string
moduli were considered to be possible inflaton candidates. 
Recently \cite{Dvali:1998pa} the
possibility that the inflaton might be an open string
modulus\footnote{Once supersymmetry is broken this open string
  mode is no longer a modulus.} has been 
investigated in a number of papers (for reviews and references see for
example
 \cite{Quevedo:2002xw,Cline:2006hu,HenryTye:2006uv,Kallosh:2007ig}). This
option is interesting 
and novel in a number of ways. One important aspect of this idea is
that 
the kinetic terms for the inflaton are
non-canonical \cite{Silverstein:2003hf,Alishahiha:2004eh} and are uniquely  
determined to all orders in $\alpha'$ 
by the Dirac-Born-Infeld action \cite{Leigh:1989jq}. It is thus a very
distinctive consequence of identifying 
the inflaton 
with an open string mode. 
The DBI kinetic terms involve higher than
quadratic powers of the time derivative of the inflaton and it is important
to determine whether these nonlinearities have practically measurable
consequences. 

Non-canonical kinetic terms of various degrees of generality have
been the focus of interest for some
time \cite{Armendariz-Picon:1999rj}-\cite{Kinney:2007ii}.  
Here however the form of the kinetic
energy density is determined by well motivated string computations. The
specific string models which have been considered in this
context involve a $D$-brane approaching an anti--$D$-brane lodged at the
bottom of a
throat \cite{Kachru:2003sx,Firouzjahi:2005dh,Chen:2004hu,Chen:2005ad,Shandera:2006ax,Kecskemeti:2006cg} 
in a warped Calabi-Yau 
compactification \cite{Giddings:2001yu}. While 
leaving many 
issues still unresolved at this point in time, it is clear that this type
of model represents a fairly 
generic situation in string theory. 
The stringy geometry at the heart of this scenario is
encoded at the effective field theory level in two objects: the usual
potential energy density $V$ and another function of the inflaton field 
denoted below by $f$. This latter object is related to the geometry of the
throat in the compactification manifold. 

It was pointed out in  \cite{Silverstein:2003hf} that the DBI action
accounts for a speed 
limit on moduli space. This arises because the kinetic energy density
depends on the time derivative of the inflaton via a factor $\gamma$, which
is unity at small velocity and grows without bound as the speed
limit is reached. 
Attention has mostly focused on the case when this limit is
attained \cite{Silverstein:2003hf,Alishahiha:2004eh,Chen:2004hu,Chen:2005ad}.
The term DBI inflation has been used mostly to refer to this case. Here this
term will be used more generally to refer to inflation in models with
Dirac-Born-Infeld kinetic terms even when far from the speed limit. 
This 
note is devoted to the more mundane situation when $\gamma$ is close to
unity, i.e. when the DBI action only gives small corrections to the
standard results. This is potentially relevant since it looks like models
with large $\gamma$ may have difficulties accommodating limits on
nongaussianity and the primordial
perturbations \cite{Shandera:2006ax,Lidsey:2006ia,Baumann:2006cd,Kecskemeti:2006cg}.

The slow-roll approximation \cite{Steinhardt:1984jj}, developed in the
context of 
effective field theory models with canonical
kinetic terms, places flatness constraints on the potential energy density
appearing in the effective field theory action. In the DBI case the action
besides the 
potential also involves the function $f$ mentioned above, so it is
interesting to ask 
what properties must this function possess 
in order that the slow roll approximation be valid. This question is
addressed here by 
studying the slow roll expansion of Liddle, Parsons and
Barrow \cite{Liddle:1994dx}, suitably 
generalized to the case of DBI kinetic terms. The lowest order in this
expansion is the slow roll approximation; requiring that corrections to
the leading order be small provides constraints on the effective action. 
In the canonical case this type of analysis implies the well-known
conditions of the ``potential slow roll'' parameters $\epsilon_V$ and
$\eta_V$. 
In the case of Dirac-Born-Infeld models it turns out that in addition to
these parameters one also needs to consider the
dimensionless quantity $fV$ which enters the slow roll expansion at 
second non-leading order (which is then the usual $\eta_V$ parameter first
appears). The generalized slow roll conditions involve the sum of $\eta_V$ 
and $\epsilon_V fV$, so barring a cancellation the standard slow roll
conditions need to be augmented by the condition $|\epsilon_V f V|\ll 1$,
which in the slow roll approximation is equivalent to $\gamma\approx 1$. 

This note is organized as follows: section \ref{canonsect} reviews 
some aspects of the case of canonical kinetic terms and 
presents a simple iterative method for deriving the slow roll expansion, 
which easily generalizes to DBI models. This generalization is presented
in section \ref{dbisect}, where some other properties of DBI models are
also discussed. 
In particular, it is shown that in contrast with the canonical case, in the
``ultra-relativistic'' limit DBI 
models do not require that the energy density be dominated by the
potential for inflation to take place. 
In section \ref{corrsect} a number
of physically relevant 
quantities related to the power spectra of primordial perturbations are
calculated, including the small corrections arising from the DBI kinetic
terms. These results are compared 
to the corresponding quantities computed in the ``ultra-relativistic''
regime.  Finally, section
\ref{concsect} offers some closing remarks.

\section{The slow roll expansion in the canonical case}
\label{canonsect}

To introduce notation, this section reviews the slow roll expansion in the
case of a canonical kinetic term. The effective action for the inflaton is
then of the form
\be
S = - \int d^4x \sqrt{-g}(\half (\d\phi)^2 + V(\phi)) \ .
\ee
For spatially homogeneous field configurations this leads to field equations
for a perfect fluid with 
\bea
p     &=&\half \dot{\phi}^2 - V(\phi) \label{canonp}\\
\rho  &=&\half \dot{\phi}^2 + V(\phi) \label{canonrho} \ .
\eea
The Einstein equations reduce to
\bea
\dot{\rho} &=& - 3 H (p+\rho) \label{conserv}\\
3\mps H^2&=& \rho \label{friedman} \ ,
\eea
where $M_P$ is the reduced Planck mass ($M_P^2=1/8\pi G$), the dot indicates
a time derivative and $H\equiv \dot{a}/a$.   

It is convenient to write these equations in first order
form \cite{Markov:1988yx,Muslimov:1990be,Salopek:1990jq,Kinney:1997ne}, 
treating  
$\phi$ as the evolution parameter in place of $t$. From \req{canonp} --
\req{friedman} it follows that  
\be
\dot{\phi} = -2 M_P^2 H'(\phi) \ ,
\ee
where the prime denotes a derivative with respect to $\phi$. Using this in
\req{friedman} gives 
\bel{hjcan}
3\mps H^2 - V = 2 M_P^4 H'^2 \ .
\ee
This is the Hamilton-Jacobi form of the field
equations \cite{Markov:1988yx,Muslimov:1990be,Salopek:1990jq,Kinney:1997ne}.   

Given a potential density $V(\phi)$ one would like to 
solve the nonlinear equation \req{hjcan} for $H(\phi)$, but this is very
difficult to do in most cases. However, if one is interested in an
inflating trajectory then there is a systematic expansion which can be
carried out analytically, in principle to any order. To formulate it one
needs to have an indication of which trajectories are inflationary. To this
end one defines  
\bel{epsilonh}
\epsilon_H = -\frac{\dot{H}}{H^2} \ ,
\ee
which satisfies
\be
\frac{\ddot{a}}{a} = H^2 (1-\epsilon_H) \ ,
\ee
so that inflation occurs if and only if $\epsilon_H<1$. In terms of
$H(\phi)$ one finds that
\be
\epsilon_H = 2\mps (\frac{H'}{H})^2 \ .
\ee

It is sometimes convenient to express the condition $\epsilon_H<1$ 
in terms of 
the ratio $\epsilon$ of kinetic energy density to
potential energy density defined by the equation:
\be
H^2 = \frac{1}{3\mps} V (1+\epsilon) \ .
\ee
Using \req{hjcan} it is easy to show that 
\be
\epsilon_H = \frac{3\epsilon}{1+\epsilon} \ .
\ee
Using this the condition for inflation reads $\epsilon<1/2$, which is 
sometimes stated as the requirement that during inflation the potential
should dominate the energy density. 
It will be shown below that in the case of DBI kinetic terms this is no
longer the case. 

The slow roll expansion of Liddle, Parsons and Barrow  \cite{Liddle:1994dx} 
(which extended the earlier work of Salopek and Bond \cite{Salopek:1990jq}) 
is based on the observation that 
an inflating trajectory has $\epsilon_H<1$, which means that the right hand
side of \req{hjcan} is small relative to the terms on the left hand side of
that equation. For this to be effective it is essential that the 
solution attains a late-time attractor \cite{Liddle:1994dx}. 
The ensuing expansion may be
derived in a number of ways. The method 
presented here has the benefit of being very simple to compute in an
iterative way. Since it is clear that the expansion is one in the number of
derivatives, one may introduce a parameter $\alpha$ which counts
derivatives by 
replacing \req{hjcan} with
\be
3\mps H^2 - V = \alpha M_P^4 H'^2 \ ,
\ee
and solving this equation in powers of $\alpha$:
\bel{hexpan}
H = \sum_n \alpha^n H_n \ .
\ee
At the end one is to set $\alpha=2$. Continuing up to second order in
$\alpha$ as above leads to: 
\bea
3\mps H_0^2 - V &=&  0\\
2 H_0 H_1 &=& \mps H_0'^2 \\
H_1^2 + 2H_0 H_2 &=& 2\mps H_1' H_0' \ .
\eea
It is trivial to write down higher orders; proceeding up the second order
as above leads to 
\bel{lpb}
H(\phi) \simeq \sqrt{\frac{V(\phi)}{3\mps}} 
\{1 + \half \epsilon_V + \epsilon_V (\eta_V-\frac{13}{8}\epsilon_V)\} \ ,
\ee
where
\beal{psrparams}
\epsilon_V &=& \frac{\mps}{6} (\frac{V'}{V})^2 \nonumber \\
\eta_V &=& \frac{\mps}{3} \frac{V''}{V} \ .
\eea
are the standard ``potential slow roll'' (PSR) parameters\footnote{The 
  normalization adopted here differs by a factor of $1/3$ from what is most  
commonly used.}. With this normalization the slow roll
expressions for the Hubble slow roll parameter and the ratio of kinetic to
potential energy are:
\be
\epsilon_H \simeq 3\epsilon_V,\quad \epsilon \simeq \epsilon_V \ .
\ee

Equation \req{lpb} coincides\footnote{Apart from the different
  normalization of $\epsilon_V$ and $\eta_V$ the formula in
   \cite{Liddle:1994dx} is 
  given for $H^2$ rather than $H$.}
with the expansion presented by Liddle, Parsons and Barrow \cite{Liddle:1994dx}. 
It shows that the validity of this expansion rests on the smallness of 
$\epsilon_V$ as well as $\eta_V$. 

A number of important calculations in the cosmology literature have been
carried out in the Hubble slow roll approximation, which uses $\epsilon_H$
and its analogues expressed in terms of higher derivatives of
$H(\phi)$. However, if one wishes to make
contact with an effective field theory defined by a lagrangian, it is
necessary to be able to calculate $H(\phi)$ in terms of the potential and
any other objects and parameters appearing in the action. 
For this purpose the expansion
\req{lpb} is very effective when applicable. The PSR parameters
\req{psrparams} are convenient, 
since they can be calculated immediately for any given model.


\section{The slow roll expansion for DBI inflation}
\label{dbisect}

The effective action for the inflaton is
the Dirac-Born-Infeld action, which for spatially homogeneous inflaton
configurations takes the form \cite{Silverstein:2003hf,Shandera:2006ax} 
\bel{dbi}
S = - \int d^4x\ a(t)^3\ \{f(\phi)^{-1}(\sqrt{1-f(\phi) \dot{\phi}^2}-1) +  
V(\phi)\}  \ .
\ee
The function $f$ appearing here can be expressed in terms of the warp
factor in the metric and the $D$3-brane tension. 
A number of specific models of this kind have
been 
studied \cite{Alishahiha:2004eh,Chen:2004hu,Chen:2005ad,Kecskemeti:2006cg,Chen:2006nt},
mostly assuming an $AdS_5$ throat ($f\sim1/\phi^4$) and various forms of
the potential $V$. While this note does not 
consider specific examples, one should keep in mind that a number of
models of this type are of interest. 

The action \req{dbi} leads to field equations \req{conserv}, \req{friedman} 
for a perfect fluid with 
\bea
p     &=& \frac{\gamma-1}{f\gamma} - V(\phi)\\
\rho  &=& \frac{\gamma-1}{f} + V(\phi) \ ,
\eea
where 
\be
\gamma = \frac{1}{\sqrt{1-f(\phi)\dot{\phi}^2}} \ .
\ee
To rewrite
the field equations in 
Hamilton-Jacobi form one proceeds as in the canonical case. From
\req{conserv}--\req{friedman} it follows that in this case that 
\bel{phidot}
\dot{\phi} = -\frac{2\mps}{\gamma} H'(\phi) \ .
\ee
This equation can easily be solved for $\dot{\phi}$ which allows one to 
express $\gamma$ as a function of $\phi$: 
\bel{gamma}
\gamma=\sqrt{1+ 4 M_P^4 f H'^2} \ .
\ee
Using this in \req{friedman} gives
\bel{hjdbi}
3\mps H^2 - V =  \frac{\gamma-1}{f} \ .
\ee
This is the Hamilton-Jacobi equation for DBI
inflation \cite{Silverstein:2003hf}. 
Solving this equation for $H(\phi)$ given $V(\phi)$ and $f(\phi)$ is rather
difficult.
Most studies have considered the
``ultra-relativistic'' case $\gamma\gg 1$, where \req{hjdbi} simplifies
considerably \cite{Silverstein:2003hf}.  
It is however clear that one can also implement the slow roll expansion
here. 

To do this, one needs to quantify under what conditions inflation takes
place for this system. As discussed in section \ref{canonsect}, in general
inflation is taking place\footnote{Of course for cosmology it is important
  that inflation takes place for long enough to generate adequate
  expansion.} whenever $\epsilon_H<1$, 
where $\epsilon_H$ is defined by eq. \req{epsilonh}. 
In the case of DBI inflation one finds (using \req{phidot}) that 
\be
\epsilon_H = \frac{2\mps}{\gamma} (\frac{H'}{H})^2 \ .
\ee
Defining the ratio $\epsilon$ of kinetic energy density to
potential energy density as before one finds in the DBI case that 
\bel{dbiepsrel}
\epsilon_H = \frac{\gamma+1}{2\gamma}\frac{3\epsilon}{1+\epsilon} \ . 
\ee
At $\gamma$ close to unity this reduces to the canonical result, but for
large gamma  
the additional factor in the above tends to $\half$. The consequence of
this is that the condition $\epsilon_H<1$ requires only $\epsilon<2$: in
this case the potential does 
not need to dominate the energy density for the system to inflate, 
unlike the case of canonical kinetic terms, when inflation requires
$\epsilon<1/2$.   

For the purpose of determining the slow roll expansion it will be
convenient to work with the square of the Hamilton-Jacobi
equation \req{hjdbisq}: 
\bel{hjdbisq}
(3\mps H^2-V) \{1 + \half f\ (3\mps H^2 - V)\} = 2 M_P^4 H'^2 \ .
\ee
One may now proceed as in the canonical case. For an inflating trajectory
on can treat the right hand side of \req{hjdbisq} as small relative to the
other terms. Introducing the 
derivative-counting parameter $\alpha$ as before and looking for the
solution in the form \req{hexpan} one obtains, to second order, 
\bea
(3\mps H_0 - V) \{1 + \half f (3\mps H_0 - V)\} &=&  0\\
2 H_0 H_1 &=& \mps H_0'^2 \\
(1+ 6 M_P^2 f H_0^2) H_1^2 + 2H_0 H_2 &=& 2\mps H_1' H_0' \ .
\eea
This yields 
\be
H(\phi) \simeq \sqrt{\frac{V(\phi)}{3\mps}} \{1 + \half \epsilon_V + 
\epsilon_V (\eta_V - \epsilon_V (\frac{13}{8} + \quart fV))\} \ . 
\ee
The second order term in the above expression shows that the applicability
of the slow roll approximation in this case requires that 
\bel{etacond}
|\eta_V - \epsilon_V (\frac{13}{8} + \quart f V)| \ll 1 \ ,
\ee
in addition to the usual $\epsilon_V<1$. While a priori it could happen 
that the two contributions in \req{etacond} partially cancel, generically
one needs 
\bel{dbisrconds}
|\eta_V| \ll 1, \quad |\epsilon_V f V| \ll 1 \ .
\ee
To understand the meaning of the dimensionless quantity $f V$, note that
the Hamilton-Jacobi equation \req{hjdbi} can be written as 
$\gamma = 1 + \epsilon f V$. If slow roll is valid one has 
$\epsilon\simeq\epsilon_V$ and so
\bel{gamsr}
\gamma \simeq 1 + \epsilon_V f V \ .
\ee
Thus the second condition in \req{dbisrconds} quantifies the expectation
that $\gamma$ has to be close to unity for the slow roll approximation to apply.

\section{Corrections to inflationary observables}
\label{corrsect}

This section looks at various physical quantities of interest in the slow 
roll approximation and presents leading corrections to these quantities
arising due to the non-canonical kinetic terms. The slow roll results are
compared with the corresponding ``ultra-relativistic'' results obtained in
the limit of large $\gamma$.

The leading effect of DBI kinetic terms are corrections of order
$\epsilon_V^2 fV$. In some models, where $\epsilon_V fV$
is small but $fV$ is large these contributions would be larger than
second order corrections of order $\epsilon_V^2$ and so on.   

The basic quantity is the number of e-folds, which is given by
\be
N = \int dt H = \sqrt{\frac{3}{2}} \int \frac{d\phi}{M_P}
\sqrt{\frac{\gamma}{\epsilon_H}} \ .
\ee
Before considering the slow roll approximation 
it is convenient to write this in 
terms of $\epsilon$ using \req{dbiepsrel}:
\bel{nfolds}
N = \frac{1}{\sqrt{3}} \int \frac{d\phi}{M_P}
\frac{\gamma}{\sqrt{\gamma+1}} \sqrt{\frac{1+\epsilon}{\epsilon}} \ .
\ee
In the case of slow roll one has \req{gamsr} which in leading order yields 
\be
N =  \int \frac{d\phi}{M_P}  \frac{1}{\sqrt{6\epsilon_V}} 
(1 + \epsilon_V + \frac{3}{4} \epsilon_V f V) \ .
\ee
The correction works to increase the number of e-folds relative to the
leading slow roll result. For comparison, at large $\gamma$ one has 
\be
\gamma\simeq\epsilon fV \gg 1 \ ,
\ee
so \req{nfolds} gives 
\be
N = \frac{1}{\sqrt{3}}\int \frac{d\phi}{M_P} \sqrt{f V(1+\epsilon)} \ .
\ee
If one assumes that potential energy dominates the energy density
(i.e. that $\epsilon\ll 1$) this formula reduces to the one given in the
literature. However, as shown earlier, one does not need $\epsilon\ll 1$ to
have inflation with DBI kinetic terms, so one may actually get more e-folds
than potential domination suggests. 

It is straightforward to write down corrections to observables related
to the primordial perturbation spectra given the results of  
Garriga and Mukhanov \cite{Garriga:1999vw}, who have calculated the spectrum
of perturbations for arbitrary kinetic terms to leading order in the
Hubble slow roll parameters $\epsilon_H$ and 
\bea
\eta_H &=& -\frac{1}{H} \frac{d}{dt} \ln \epsilon_H \label{heta}\\
\sigma_H &=& -\frac{1}{H} \frac{d}{dt} \ln c_s \label{hsigma}\ .
\eea
The required smallness of $\eta_H $ and $\sigma_H$ expresses the condition
that $\epsilon_H$ and the speed of sound $c_s$ should vary much more slowly
than the scale factor. The results of \cite{Garriga:1999vw} used below
assume that $\epsilon_H$, $\eta_H $, $\sigma_H$ are small during the
observable phase of inflation. Use of these formulae in the
``ultra-relativistic'' regime is also valid only under this
assumption\footnote{An interesting example where these formulae break down
  as a consequence of the slow roll conditions being violated is
  discussed in reference \cite{Kinney:2007ii}.}.  

The result for the primordial scalar power spectrum is
\cite{Garriga:1999vw} 
\be
P_{\cal R} = \frac{1}{36 \pi^2 M_P^4} \frac{1}{c_s} \frac{\rho^2}{p+\rho} \ ,
\ee
For DBI models one
has \cite{Alishahiha:2004eh} $c_s=1/\gamma$. Here and in all the formulae
below the right 
hand side is evaluated at the sound horizon exit ($c_s k = a H$). 
In the slow roll expansion one finds
\bel{rpower}
P_{\cal R} \simeq \frac{1}{72\pi^2  M_P^4} \frac{V}{\epsilon_V} (1 +
\epsilon_V f V)  \ ,
\ee
while in the ``ultra-relativistic'' limit 
\be
P_{\cal R} \simeq \frac{1}{36\pi^2  M_P^4} (1+\epsilon)^2 f V^2 \ .
\ee
In models where $\epsilon\ll 1$ this is the same as found in
 \cite{Alishahiha:2004eh,Chen:2006nt}.  

For the tensor perturbations one has \cite{Garriga:1999vw}
\bel{hpower}
P_h = \frac{2}{\pi^2 M_P^2} H^2 \ .
\ee
This gives 
\be
r\equiv \frac{P_h}{P_{\cal R}} = \frac{16\epsilon_H}{\gamma} \ ,
\ee
which in the slow roll approximation yields 
\be
r \simeq 48\epsilon_V (1 - \epsilon_V f V) \ .
\ee

For the spectral indices the results of Garriga and
Mukhanov \cite{Garriga:1999vw} can be written as
follows:
\beal{indices}
n_s-1&=&-2\epsilon_H + \eta_H + \sigma_H\\
n_T &=& -2 \epsilon_H \ .
\eea
These expressions are valid in the leading order in Hubble slow roll
parameters. 
In the Hamilton-Jacobi formalism one finds 
\bea
\eta_H &=& \frac{4}{3\gamma}\frac{H''}{H} - 2\epsilon_H + \sigma_H\\
\sigma_H &=& -\frac{2}{3}\frac{H'}{H}\frac{\gamma'}{\gamma} \ ,
\eea
which leads to the expression 
\be
n_s-1=-4\epsilon_H + 2\sigma_H + \frac{4}{3\gamma} \frac{H''}{H} \ .
\ee
Evaluating this to leading order in the slow roll expansion
gives\footnote{The unusual coefficients in this formula are a consequence
  of the adopted normalization of the potential slow roll parameters.}
\be
n_s - 1 = -18\epsilon_V + 6\eta_V + \frac{8}{27} \epsilon_V f' V' \ . 
\ee
The correction due to the DBI kinetic terms can be of either sign depending
on the model.

\section{Conclusions}
\label{concsect}

Spacetime-filling $D$-branes moving on the $6$-dimensional compact space
appearing in a string compactification is
a very natural 
setting for inflation in string theory. While many structural aspects of
this scenario are already known, many details which will be essential in
determining whether this is how it happened are still unknown. Apart from
questions about the precise ending of inflation, brane annihilation, energy
transfer to standard model degrees of freedom and so on, there are still
many technical aspects which need to be understood. This includes a
reliable computation of the potential and the geometry of the
throat in which the brane is moving while spacetime inflates. 
Existing examples are probably just the beginning. 
It is not 
even completely clear whether all the $D$-brane moduli effectively reduce
to a single one relevant for driving inflation, as assumed in the class of
effective field theories discussed here. While precise studies will
clearly require numerical methods it is essential to explore approximate
analytic tools in the regimes where some are available. 

The main focus of this note was to look at the validity of the slow roll
approximation for effective inflaton field theories with Dirac-Born-Infeld
kinetic terms, which are are a rather distinctive feature of brane
inflation. The slow roll expansion introduced in the case of canonical
kinetic terms has a natural generalization to DBI models which can be
computed easily using the iterative method presented here. It is obviously
worthwhile to study such corrections in specific instances.

In the other extreme, that of large $\gamma$, is interesting that DBI
models allow inflation with a higher ratio of kinetic to potential energy
than is possible with canonical kinetic terms. This could significantly
affect predictions at least in some cases.

\end{document}